\documentclass[aps,prb,preprint,groupedaddress,showpacs]{revtex4}
\usepackage{graphicx}
\usepackage{amsmath}
\usepackage{amssymb}
\usepackage{color}
\usepackage{bm}       
\usepackage{flafter}
\def\br{ \bm{r} }

\def\re{ \,\mathrm{Re}\,}

\begin{document}


\title{Quasiparticle-Self-Energy and its Effect on the Superconducting Order Parameter  of the Pyrochlore Superconductor Cd$_2$Re$_2$O$_7$ }

\date{\today}

\author{F. S. Razavi}

\affiliation{Department of Physics, Brock University, \mbox{St. Catharines}, Ontario, L2S 3A1, Canada}

\author{Y. Rohanizadegan}
\affiliation{Department of Physics, Brock University, \mbox{St. Catharines}, Ontario, L2S 3A1, Canada}

\author{M. Hajialamdari}
\affiliation{Department of Physics, Brock University, \mbox{St. Catharines}, Ontario, L2S 3A1, Canada}

\author{M. Reedyk}
\affiliation{Department of Physics, Brock University, \mbox{St. Catharines}, Ontario, L2S 3A1, Canada}

\author{B. Mitrovi\'c}
\affiliation{Department of Physics, Brock University, \mbox{St. Catharines}, Ontario, L2S 3A1, Canada}

\author{R. K. Kremer}
\affiliation{Max-Planck-Institut f\"{u}r Festk\"{o}rperforschung,
Heisenbergstra$\rm\beta$e 1, 70569 Stuttgart, Germany}

\date{\today}

\begin{abstract}
The magnitude and the temperature dependence of the superconducting order parameter $\Delta(T)$ of
single-crystals of Cd$_2$Re$_2$O$_7$ ($T_c$ = 1.02~K) was measured using point
contact spectroscopy. In order to fit the conductance spectra and to extract the order parameter
at different temperatures we generalized
the Blonder-Tinkham-Klapwijk theory by including the self-energy of the quasiparticles into the Bogoliubov  equations.
This modification enabled excellent fits of the conductance spectra.
$\Delta (T)$  increases steeply below the superconducting transition temperature of 1.02 K
and levels off below $\sim$0.8 K depending on measurement directions where its value varried from 0.22(1) meV to 0.26(1) meV.
Our results indicate the presence of a strong electron-phonon interaction and an enhanced quasiparticle damping
and may be related to a possible phase transition within the superconducting region at $\sim$0.8 K.

\end{abstract}

\maketitle

A large number of transition metal (TM) oxides crystallize with the $\alpha$-pyrochlore structure with the general
composition A$_2$B$_2$O$_7$, where A is a larger and B is a smaller TM cation. Amongst the $\alpha$-pyrochlores,
Cd$_2$Re$_2$O$_7$ is the only one which shows superconductivity at $\sim$1 K \cite{Hanawa,sakai}.
At room temperature, Cd$_2$Re$_2$O$_7$ exhibits a cubic structure (space group $Fd\bar{\rm 3}m$).
At $T_{\rm S1} \sim$ 200 K, Cd$_2$Re$_2$O$_7$ undergoes a metal-to-metal second order structural
phase transition (PT) to a non-centrosymmetric tetragonal structure (space group $I\bar{\rm 4}m$2) followed
by a first order PT at $T_{\rm S2} \sim$ 120 K to another tetragonal structure (space group $I$4$_1$22)
\cite{Hanawa2,jin,arai}. These two PTs have a profound effect on the electronic and the magnetic properties
of Cd$_2$Re$_2$O$_7$. The electrical resistivity and the magnetic susceptibility drop sharply below
$T_{\rm S1}$ \cite{Hanawa2,jin}. Heat capacity measurements below $T_{\rm S2}$ showed an exceptionally
large electronic Sommerfeld coefficient of $\gamma $ =15 mJ/K$^2$ mol Re\cite{Hanawa,hiroi}. Band structure
calculations for the room-temperature cubic structure revealed that the electronic density of states (DOS)
at the Fermi level arises mainly from bands with Re-5$d$ character  with electron or hole pockets at various
points of the Brillouin zone \cite{singh}. However, the  band structure of Cd$_2$Re$_2$O$_7$ in the
low-temperature structure ($T < T_{\rm S2}$)  indicated localized Cd 4$d$ and itinerant Re-5$d$
electrons and a quasi two dimensional Fermi surface \cite{Huang}. The results of  the Re nuclear
quadrupole resonance (NQR) and the Cd nuclear magnetic resonance (NMR) at low temperature, and in
the superconducting state, revealed no magnetic or charge ordering \cite{vyaselev}.
Just below $T_c$, the $^{\rm 187}$Re spin lattice relaxation rate exhibits a pronounced
coherence peak  with an increase of the relaxation rate by a factor of two and subsequently,
below $\sim$0.8~K, an exponential decrease consistent with weak-coupling BCS theory and an isotropic
gap \cite{vyaselev}. Vyaselev \textit{et al.} calculated the Wilson ratio and obtained a value of 0.34
indicating strong electron-phonon coupling incompatible with weak-coupling theory \cite{vyaselev}.
The far-infrared spectroscopy measurements on Cd$_2$Re$_2$O$_7$ crystals in the superconducting
state at $\sim$0.5~K  revealed two strong absorption peaks near 9.6 and 19.3 cm$^{-1}$ which completely
vanish above $T_c$ possibly indicating strong electron-phonon coupling
in the superconducting state \cite{Hajialam}.

In this article, we report point contact spectroscopy measurements on single crystals of Cd$_2$Re$_2$O$_7$
carried out in order to  better understand the superconducting state of Cd$_2$Re$_2$O$_7$. Our measurements
provide the temperature dependence of the superconducting order parameter which differs markedly from the
BCS temperature dependence of the energy gap. For $T \rightarrow$ 0~K the order parameter approaches a value
between 0.22(1)to 0.26(1) meV, i.e 2$\Delta(0)$/$k_{\rm B}T_c >$ 5.0(1) indicating that Cd$_2$Re$_2$O$_7$ is a
strong-coupling superconductor.

\begin{centering}
\begin{figure}[h]
\includegraphics[width=8cm]{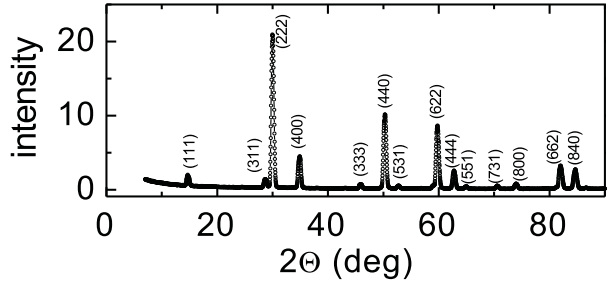}
\caption{ Powder X-ray diffraction result of crushed single crystals
of Cd$_2$Re$_2$O$_7$.} \label{Fig1}
\end{figure}
\end{centering}

Single crystals of Cd$_2$Re$_2$O$_7$ were grown from Re$_2$O$_7$ and Cd (purity 99.99\%) by a gas phase chemical transport technique \cite{Barisic}.
Re$_2$O$_7$ powder was prepared  from Re metal (purity 99.99\%) and O$_2$
(purity 99.9999\%) by employing the Noddack  method \cite{Noddack}.
To check the purity of the samples, we crushed several small crystals and performed  X-ray powder diffraction measurements ( see Fig.~\ref{Fig1}) where we found no impurity phases. The
lattice parameter was refined to 1.028(5) nm, very close to the published data \cite{Hanawa}.
Electron-microprobe analysis of selected Cd$_2$Re$_2$O$_7$ crystal proved an ideal Cd:Re ratio of one, within error bars of the method.

\begin{centering}
\begin{figure}[h]
\includegraphics[width=8cm]{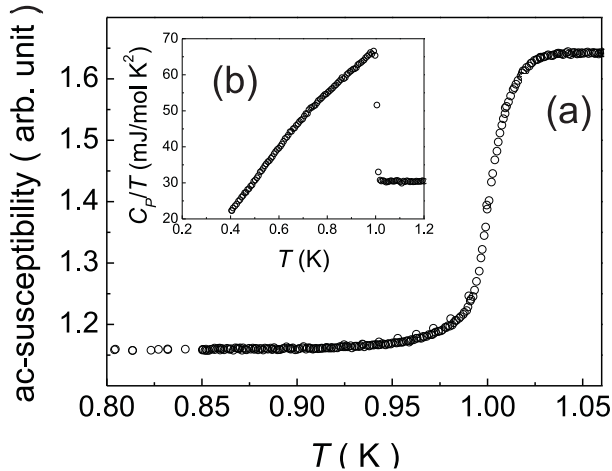}
\caption{(a) Temperature dependence of the ac-susceptibility  of a single
crystal of Cd$_2$Re$_2$O$_7$. (b) Heat capacity of Cd$_2$Re$_2$O$_7$.}
\label{Fig2}
\end{figure}
\end{centering}

 The ac-susceptibility and the heat-capacity measurements
showed a superconducting transition at 1.02 K and a width of the transition of $\pm$30 mK
(see Fig.~\ref{Fig2}), consistent with data reported previously \cite{hiroi}.

Andreev reflection spectra were measured in a specific crystallographic plane by employing the soft point
contact spectroscopy (PCS) method reviewed  in detail by Daghero and Gonnelli
 \cite{Gonnelli}. The resistance of the contact to
the sample was adjusted to be within the range (of $\sim$10 Ohms) of
the resistance of the sample by repeatedly applying short high
voltage pulses through the contacts\cite{daghero}. The sample was
cooled in a home-built single-shot $^3$He cryostat which enables the
samples to be fully immersed in the cryogenic fluid reducing ohmic
heating of the point contacts. The temperature was stabilized by
controlling vapor pressure of liquid $^3$He and it was measured
 with a custom calibrated Cernox CX1030 temperature sensor.
 \begin{centering}
\begin{figure}[h]
\includegraphics[width=8cm]{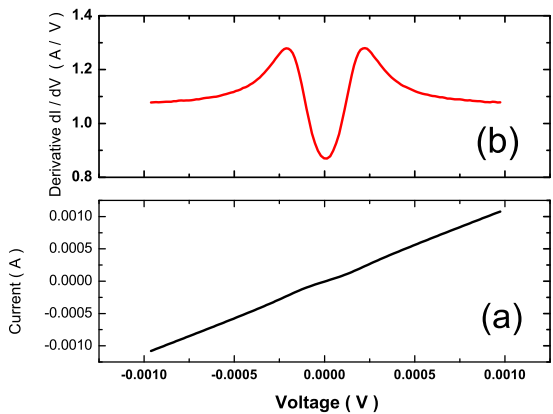}
\caption{(a) A sample of I vs. V and (b) the calculated $dI \over
dV$; no averaging was used in the process. The maximum error  on
the derivative is of the order of $\pm 0.002$ and can not be represented
on the graph.} \label{Fig3}
\end{figure}
\end{centering}

 The stability of temperature measurements was less than $\pm$3 mK below
and $\pm$7 mK above 0.7 K, respectively. A graph of raw data I vs V is
shown in Fig.~\ref{Fig3}a and its derivative is shown in  Fig.~\ref{Fig3}b.  No data
averaging was done post  measurement. The accuracy in current
measurements is of the order of $ 10^{-9}$ A and for voltage of the
order of $\pm 2\times 10^{-8}$ V. The  measurement error on $dI\over
dV$ is of the order of 0.2\%  of its value. The fitting program stops iteration when the
total standard deviation of all points is less than $10^{-8}$.

\begin{centering}
\begin{figure}
\includegraphics[width=8.5cm,height=7cm]{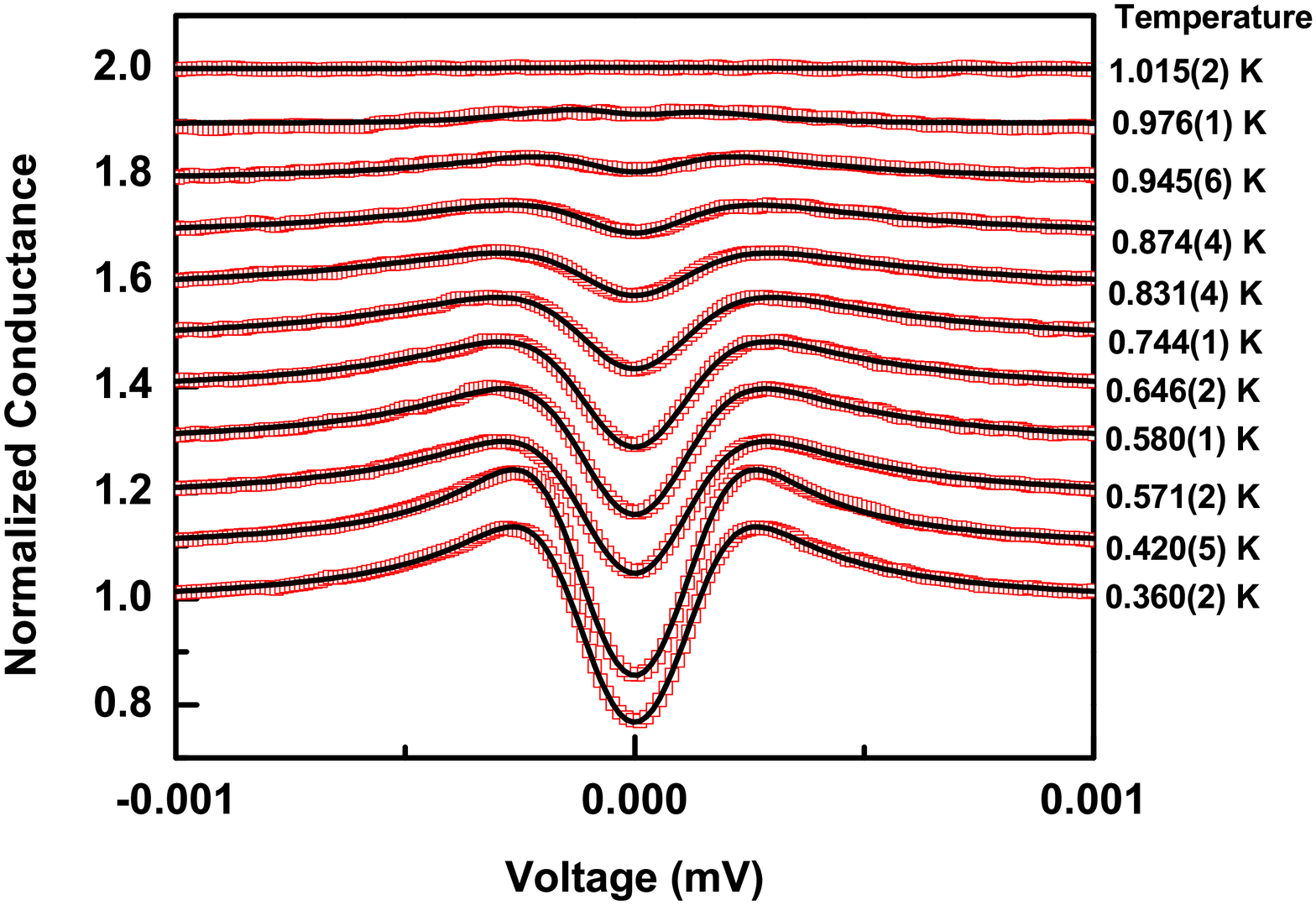}
\caption{Voltage dependence of the normalized conductance ($\Box$) for Cd$_2$Re$_2$O$_7$. For clarity the
normalized conductance is shifted by 0.1  relative to the previous temperature indicated on the right.
 The solid black lines represent the fits obtained using our generalized BTK theory with a complex
gap.}
\label{Fig4}
\end{figure}
\end{centering}

Several measurements were done by injecting current in the (001), (100), (110), and (111) planes.
As an example, a set of characteristic normalized conductance spectra collected by injecting the current in the (111) plane
is displayed in Fig.~\ref{Fig4}. All spectra were normalized to the value of the
conductance $dI/dV$ measured at a voltage $\simeq$ $\pm$2.0~mV in the normal state.

An attempt to fit the normalized conductance spectra of
Cd$_{2}$Re$_{2}$O$_{7}$ using the Blonder--Tinkham--Klapwijk (BTK)
theory \cite{BTK} did not provide satisfactory results, especially
for the spectra at temperatures  close to $T_{c}$ where
finite lifetime effects can play an important role \cite{Kaplan}.
The BTK theory is based on the Bogoliubov equations for the
two-component wave function
\begin{equation}
\label{psi}  \psi(\br,t)=\left( \begin{array}{c}
                  u(\br,t) \\
                  v(\br,t)
                   \end{array} \right)
\end{equation}
for particles and holes, respectively, but does not take into consideration any self-energy
effects, i.e.~finite quasiparticle lifetimes.
Previous attempts to extend the BTK theory to include
lifetime effects were purely phenomenological and assumed
that the time-dependence of the particle and hole
amplitudes $u$ and $v$ in the Bogoliubov equations are of the form
$\exp[-i(E-i\Gamma)t/\hbar]$, where $E$ is the quasiparticle energy,
and $\Gamma$ is its scattering rate \cite{Plecenik,deWilde,Janson}.
The resulting theory has a form identical to the BTK theory but
with the normalized superconducting quasiparticle density of states
given by the so-called Dynes equation \cite{Dynes}
$N_s(E) = \re\{{(E-i\Gamma)}/{\sqrt{(E-i\Gamma)^{2}-\Delta^{2}}}\}$,
where $\Delta$ is the superconducting energy gap. However, it was
pointed out that the above equation  cannot be
justified microscopically, at least not for conventional strong-coupling
superconductors such as Pb or Nb \cite{Mitrovic}.
In the strong-coupling (Eliashberg) theory of
superconductivity the normalized
superconducting quasiparticle density of states is given by \cite{Schrieffer,Scalapino}
\begin{equation}
\label{DOS} N_{s}(E)=\re\frac{E}{\sqrt{E^{2}-\Delta^{2}(E)}}\>,
\end{equation}
where $\Delta(E)$ is the complex gap function, i.e. the renormalized
pairing self-energy. All damping and retardation effects are
contained in $\Delta(E)$.

We have generalized the BTK theory by using the Bogoliubov equations
which include the self-energy effects \cite{McMillan}. Their
time Fourier transform has the form
\begin{equation}
\label{Bogoliubov}
\{[-\frac{\hbar^{2}}{2m}\nabla^{2}-\mu(\br)]\tau_{3}+\Sigma(\br,E)\}\psi(\br,E)=E\psi(\br,E),
\end{equation}
where
\begin{equation}
\label{self-energy}
\Sigma(\br,E)=(1-Z(\br,E))\tau_{0}+\phi(\br,E)\tau_{1}
\end{equation}
is the 2$\times$2 electron self-energy matrix. $\tau_{0}$ is a unit
matrix and $\tau_{1}$ and $\tau_{3}$ are Pauli matrices
\cite{Schrieffer,Scalapino}. We have assumed that the self-energy is local
in space which is justified if it arises from the
electron-phonon interaction. The gap function $\Delta(\br,E)$ is
related to the pairing self-energy $\phi(\br,E)$ and the
renormalization function $Z(\br,E)$ by
$\Delta(\br,E)=\phi(\br,E)/Z(\br,E)$. In the weak-coupling
limit $Z=$1, $\Delta=\phi$ and Eq. (\ref{Bogoliubov}) reduces to the
familiar Bogoliubov equation.

Subsequently, by making the same assumptions as in the derivation of the BTK theory
(spatially independent $\mu$, $\phi$ and $Z$,
translational invariance along $y$- and $z$-directions and a
$\delta$-function potential at the normal metal (N)--superconductor
(S) interface) \cite{BTK}, we arrive at a theory which is identical in form with
the BTK theory \cite{BTK} but with the real gap $\Delta$
replaced by the complex gap function $\Delta(E)$. Details of the
derivation are published elsewhere \cite{yousef}. Specifically,
the conductance of an N--S interface at a voltage $V$ is given by
\begin{equation}
\label{conductance}
\frac{dI_{NS}}{dV}=S\int_{-\infty}^{+\infty}dE\frac{df(E-eV)}{dV}[1+A(E)-B(E)]\>,
\end{equation}
with $S=2N(0)ev_{F}\cal{A}$. $N(0)$ is the single-spin Fermi
level density of states in the normal state, $e$ is the electron
charge, $v_{F}$ is the Fermi velocity and $\cal{A}$ the effective
area of the N--S interface. The probability current densities for
the Andreev reflection $A(E)$ and for the normal reflection $B(E)$
are given by (in units of the Fermi velocity $v_{F}$)
\begin{eqnarray}
\label{details1}
A(E) & = & \frac{|u|^{2}|v|^{2}}{|\gamma|^{2}} \\
\label{details2}
B(E) & = & \frac{[|u|^{4}+|v|^{4}-2\re(u^{2}v^{2})]z^{2}(z^{2}+1)}{|\gamma|^{2}} \\
\label{details3}
\gamma & = & u^{2}+(u^{2}-v^{2})z^{2} \\
\label{details4}
u, v & = &
\frac{1}{\sqrt{2}}\sqrt{1\pm\sqrt{E^{2}-\Delta^{2}(E)}/E}\>.
\end{eqnarray}
The parameter $z$ in Eqs.(\ref{details2}-\ref{details3}) is a
dimensionless barrier strength parameter related to the strength of
the $\delta$-function potential $V(x)=H\delta(x)$ at the interface
by $z=H/(\hbar v_{F})$.
\begin{centering}
\begin{figure}
\includegraphics[width=8.5cm,height=7cm]{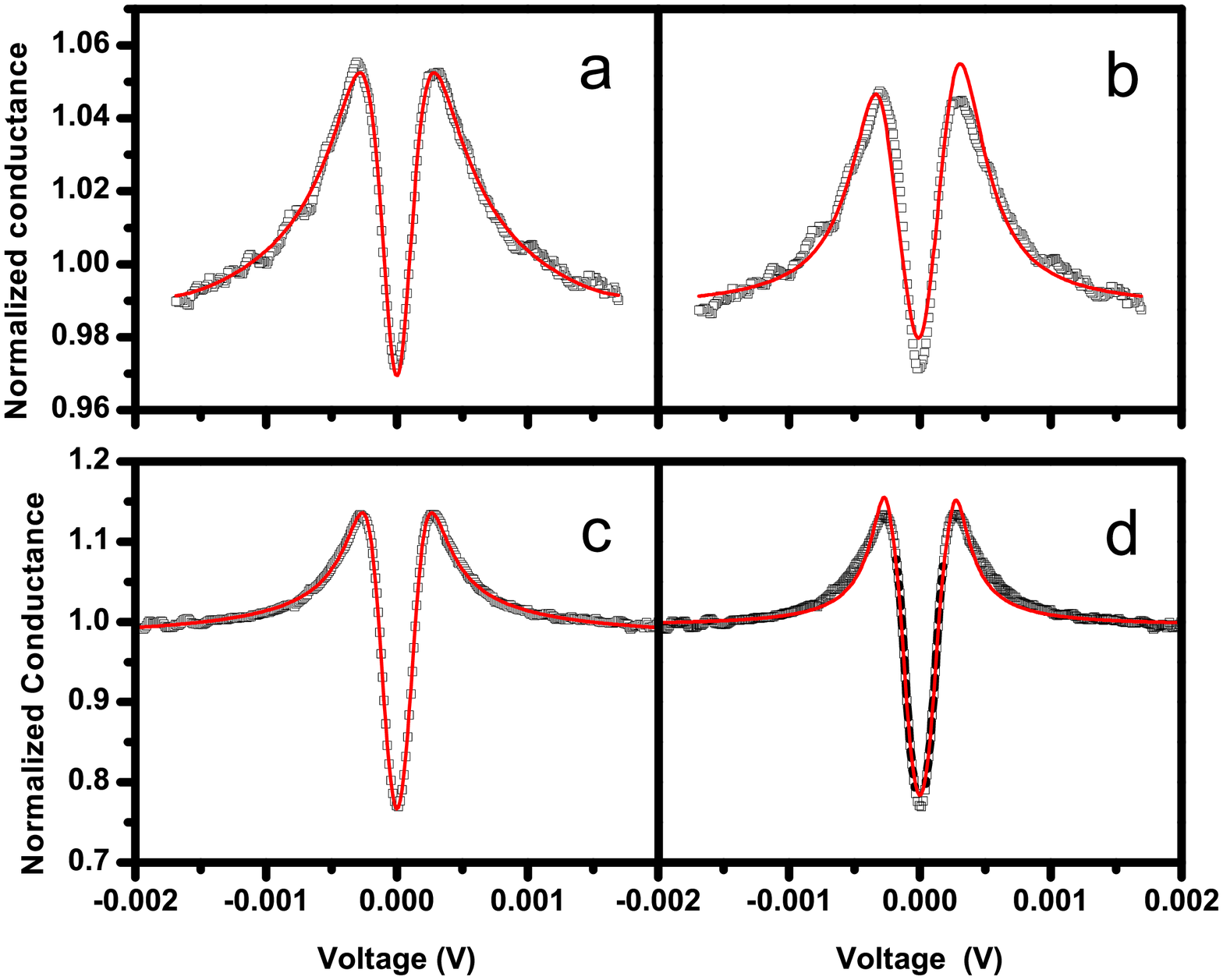}
\caption{Experimental data ($\Box$) and theoretical fits for the normalized conductance spectra
of Cd$_{2}$Re$_{2}$O$_{7}$ at 0.831 K (a and b) and  0.360 K
(c and d). Solid lines in Figs. \ref{Fig5}a and \ref{Fig5}c represent the fits obtained using our
generalized BTK theory with a complex
gap. Solid lines in Figs. \ref{Fig5}b and \ref{Fig5}d show the fits using the phenomenological
approach including the Dynes broadening parameter $\Gamma$ \cite{Plecenik,deWilde,Janson}.
}
\label{Fig5}
\end{figure}
\end{centering}

\begin{centering}
\begin{figure}[h]
\includegraphics[width=8cm]{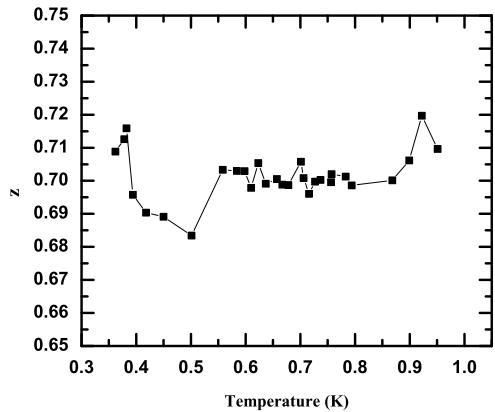}
\caption{Variation of z as a function of temperature with a total
random fluctuation of  $\pm $ 3\%. } \label{Fig6}
\end{figure}
\end{centering}

For $E$ not too far from the gap edge the real and the imaginary
part of $\Delta(E)$ can be taken as constant \cite{Kaplan,Mitrovic}.
Thus, in applying Eqs.(\ref{conductance}-\ref{details4}) to the
experimental results there are three fit parameters at each
temperature: the real and the imaginary part of the gap at the gap
edge, and the barrier strength parameter $z$.  Two characteristic
fits to the conductance spectra ($dI/dV$ vs $T$) at different
temperatures are displayed in Fig.~\ref{Fig5}a, and
\ref{Fig5}c. We found that the
theoretical fit using the  BTK model modified by including a complex
$\Delta$ provides a significantly improved agreement with the
experimental data.   The
variation of z parameter as a function of temperature did not show
any systematic variation with temperature (see Fig.~\ref{Fig6}) and it shows a random
fluctuation with temperature of the order of $\pm $ 3\%.
In contrast, the fits with the BTK model extended by including a
phenomenological broadening parameter $\Gamma$ (Fig.~\ref{Fig5}b, and
\ref{Fig5}d) describe the experimental data less well, especially
for temperatures near $T_c$. This finding  renders strong support
that the described theoretical amendments are essential for a
description of conductance spectra of superconductors in the strong
coupling limit.
\begin{centering}
\begin{figure}
\includegraphics[width=8.5cm,height=7cm]{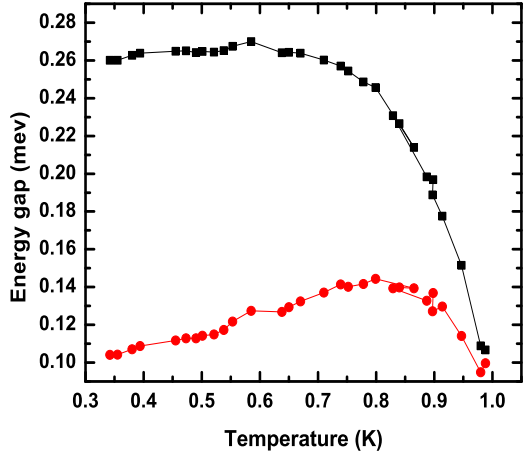}
\caption{Superconducting energy gap $\Delta_{0}(T)$ of Cd$_{2}$Re$_{2}$O$_{7}$ (square symbols)for (001) plane obtained by
fitting the experimental data for the normalized conductance at different temperatures using  the generalized
BTK theory presented in this work. The circles
  give the values of the imaginary part of the gap at the gap edge obtained in the fits (the errors
  in fitted values are less than the size of the symbols).}
\label{Fig7}
\end{figure}
\end{centering}

\begin{centering}
\begin{figure}
\includegraphics[width=8.5cm,height=7cm]{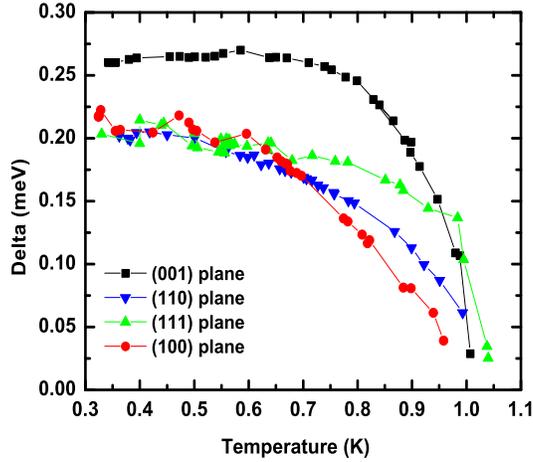}
\caption{$\Delta_{0}(T)$ of Cd$_2$Re$_2$O$_7$ measured for different crystallographic planes}
\label{Fig8}
\end{figure}
\end{centering}

By fitting the set of temperature dependent conductance spectra, we obtained the superconducting
energy gap of Cd$_2$Re$_2$O$_7$ as a function of temperature (see Fig.~\ref{Fig7}).
We also checked for anisotropy in  $\Delta_{0}(T)$  for different crystallographical planes as shown in
 Fig.~\ref{Fig8}. Both injected and measured voltage are in the same plane. We observed the $\Delta_{0}(T)$ values
  for all planes saturate at about  $\Delta_{0}(T)$ = 0.22 meV,except however, for the (001) plane where the value is slightly
  larger and saturated at about 0.26 meV.
The deduced values for $2\Delta(T\rightarrow0)/k_{\rm B}T_c$ which vary between 5.0(0.1)to 6.2(0.1) for different crystallographic planes
indicate a magnitude characteristic for strong-coupling superconductors \cite{mitrovic84}.
The temperature dependence of the gap is markedly different from that expected from BCS theory.
$\Delta(T)$  rises rapidly below $T_c$ and levels off below $\sim$0.8~K whereas the imaginary part exhibits a
maximum at $\sim$0.8 K and subsequently decreases again but remains constant to the lowest temperatures measured.

The magnitude and the temperature dependence of the superconducting order parameter found in this study give
an indication that Cd$_2$Re$_2$O$_7$ is a strong coupled superconductor with a remarkably large gap resulting
in a ratio 2$\Delta(0)$/$k_{\rm B}T_c >$ 5, which is comparable to values found e.g. in Pb--Bi alloys with
extremely strong electron-phonon coupling \cite{mitrovic84}. The related ternary $\beta$-pyrochlore osmium
oxides, AOs$_2$O$_6$, where A is an alkali metal cation, also exhibit such large 2$\Delta(0)$/$k_{\rm B}T_c$
ratios, as observed for KOs$_2$O$_6$ \cite{shimo}. However, $T_c$ of KOs$_2$O$_6$ is by an order of magnitude higher
than in Cd$_2$Re$_2$O$_7$ \cite{hiroi2}.
In  the superconducting region the density of states of KOs$_2$O$_6$ was measured\cite{shimo} using photoemission
spectroscopy and the data were fitted using the Dynes formula \cite{Dynes}. The resulting broadening parameter
$\Gamma$ had temperature dependence quite similar to what we found for the imaginary part of the gap shown in
Fig.~\ref{Fig7}.
An additional similarity with the $\beta$-pyrochlore osmium
superconductors is the substantial enhancement of the experimental Sommerfeld $\gamma$-term with respect
to the band $\gamma$-term. Superconductivity in the $\beta$-pyrochlore osmium oxides is usually attributed
to a low-energy rattling vibrational mode of the alkali metal atoms which may also account for the observed
first order structural phase transition occurring below $T_C$ in KOs$_2$O$_6$ \cite{shimo,hiroi2}. It is
tempting to associate the superconductivity in Cd$_2$Re$_2$O$_7$ to similar low-energy vibrational excitations,
e.g.~to the low-energy IR  modes recently observed by low-temperature optical spectroscopy
in the superconducting state\cite{Hajialam}.

\begin{centering}
\begin{figure}[h]
\includegraphics[width=8cm]{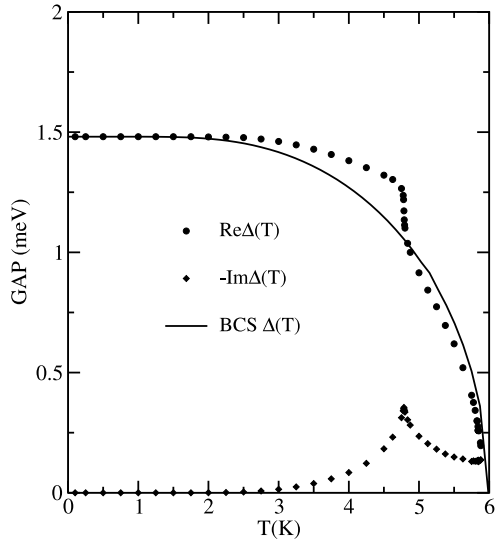}
\caption{The real and imaginary part of the gap at the gap edge as a
function of temperature obtained for a model electron-phonon
spectrum in the form of a very sharp cutoff Lorentzian at 2.2 meV
such that the electron-phonon mass renormalization parameter is
$\lambda_{R}$=3 and $k_{B}T_{c}/\omega_{\ln}$=0.24. The calculated
value of 2$\Delta_{0}(0)/(k_{B}T_{c})$ is 5.73.}
\label{Fig9}
\end{figure}
\end{centering}

 The temperature dependence of the gap edge is
quite unusual even for strongly coupled electron-phonon
superconductors such as Pb where $\Delta_{0}(T)$ closely follows the
BCS curve (see Fig.~44 in Ref. 23). In
Cd$_{2}$Re$_{2}$O$_{7}$ $\Delta_{0}(T)$ remains flat up to about
80\% of the $T_{c}$ and then it drops precipitously to 0 at $T_{c}$.
The imaginary part of the gap at the gap edge has a sharp peak near
the temperature where the gap edge starts its rapid decrease. Such
temperature dependence can result in systems where the
superconductivity is caused by electrons coupling to a low frequency
phonon mode as illustrated in Fig.~\ref{Fig9}.

The results shown in Fig.~\ref{Fig9} were obtained \cite{Mitrovic and Nicol}
for a model for the $\beta$-pyrochlore KOs$_{2}$O$_{6}$
\cite{Shimojima} where the superconductivity is assumed to result
from the electron coupling to a low frequency anharmonic (rattling)
mode \cite{Mahan}. Using a very sharp cutoff Lorentzian model for
the electron-phonon coupling function of a ``rattler''
$\alpha^{2}F_{R}(\Omega)=(g\epsilon/\pi)\left\{1/\left[(\Omega-\Omega_{R})^{2}+\epsilon^{2}\right]-
1/(\Omega_{c}^{2}+\epsilon^{2})\right\}$, for
$\Omega_{R}-\Omega_{c}\leq\Omega\leq\Omega_{R}+\Omega_{c}$, and
$\alpha^{2}F_{R}(\Omega)=$ 0, for $|\Omega-\Omega_{R}|>\Omega_{c}$,
with the rattling frequency $\Omega_{R}=$ 2.2 meV, the cutoff frequency $\Omega_{c}=$ 2.1 meV, the peak-width
parameter $\epsilon=$ 0.01 meV, and the the coupling strength parameter $g=$ 3.312 (chosen such that
$\lambda_{R}=\int_{0}^{+\infty}d\Omega\alpha^{2}F_{R}(\Omega)/\Omega=$ 3), the Eliashberg equations were solved
on the real axis at various temperatures. The real and the imaginary part of the gap, at the gap edge, defined by
$\re \Delta(E=\Delta_{0}(T),T)=\Delta_{0}(T)$, are shown in Fig.~\ref{Fig9}. The transition temperature $T_{c}$ was
obtained by solving the Eliashberg equations on the imaginary axis with the cutoff in the Matsubara sums
$\omega_{c}=$ 30 meV and the Coulomb repulsion parameter $\mu^{*}(\omega_{c})=$ 0.1.
The same $\omega_{c}$ and $\mu^{*}(\omega_{c})$ were used in
the real axis calculations. The resulting $T_{c}$ was 6 K and 2$\Delta_{0}(0)/(k_{B}T_{c})=$ 5.73.

Strong electron-phonon coupling may lead to  structural PTs as it has been observed for KOs$_2$O$_6$
below $T_c$ and attributed to a freezing of the rattling motion of the K atoms \cite{hiroi2}. A detailed
examination of our (see Fig.~\ref{Fig2}) and published heat capacity and upper critical field data of
Cd$_2$Re$_2$O$_7$ reveals  a change in the slope of these data at $\sim$0.8 K which could in fact  be due to a
freezing of low-energy lattice degrees of freedom \cite{Hanawa2}. We rule out Cd or combined Cd--O related
vibrations since preliminary results on $^{116}$Cd and $^{18}$O isotope enriched samples,
$^{116}$Cd$_2$Re$_2^{18}$O$_7$ and Cd$_2$Re$_2^{18}$O$_7$, showed no isotope effect on $T_c$ \cite{razavi2013}.
This finding rather suggests the importance of Re related lattice vibrations, possibly of a low-energy Re-lattice
breathing mode as discussed by Hanawa \textit{et al.} in context with the freezing at the high-temperature
structural PT \cite{Hanawa2}. Interestingly, ReO$_3$ also exhibits unusually large anisotropic thermal
vibrations of the oxygen atoms and a proximity to a low-pressure  structural phase transition \cite{Wdowik2010}.
 The observation of anisotropy in the superconducting order parameter  $\Delta_{0}(T)$ in Cd$_2$Re$_2$O$_7$  might be as a result of the
   anisotropy of the Fermi surface as reported by band structure calculations\cite{singh}.

In summary, we have measured N--S conductance spectra below the superconducting transition temperature
of the $\alpha$-pyrochlore superconductor Cd$_2$Re$_2$O$_7$ by soft point contact spectroscopy.
We developed and employed an
extension of the BTK theory by including the quasiparticle self-energy into the Bogoliubov equations, and
thus we were able to fit the conductance spectra as well as derive the temperature dependence and the
magnitude of the superconducting order parameter. The magnitude of the gap at $T$ = 0 indicates that
Cd$_2$Re$_2$O$_7$ is a strong-coupling superconductor. The temperature dependence of the order parameter is
markedly different from that of the weak-coupling BCS gap.

\acknowledgments
Financial support for this work was
partially provided by the Natural Sciences and Engineering Research
Council of Canada (NSERC) and the Canadian Foundation for Innovation (CFI). F.S.R. acknowledges
the research support by the MPG during his sabbatical stay at the MPI for Solid State Research,
Stuttgart, Germany.

\end{document}